\documentclass[twocolumn]{aastex631}
\usepackage{rotating}
\usepackage{graphicx}
\usepackage{supertabular}
\usepackage{longtable}
\usepackage{CJK}
\usepackage{tikz}
\usepackage{threeparttable}
\usetikzlibrary{shapes,arrows}

\received{----}
\revised{----}
\accepted{April 09, 2024}
\submitjournal{ApJS}

\begin{document}
\begin{CJK*}{UTF8}{gbsn}
\title{Searching for short-period variables in M\,31\,: method and catalogs}

\correspondingauthor{Haibo Yuan, Subo Dong}
\email{yuanhb@bnu.edu.cn, dongsubo@pku.edu.cn}

\author[0009-0007-5610-6495]{Hongrui Gu(顾弘睿)}
\affiliation{Institute for Frontiers in Astronomy and Astrophysics,
Beijing Normal University, Beijing, 102206, People's Republic of China; }
\affiliation{CAS Key Laboratory of Optical Astronomy, National Astronomical Observatories, Chinese Academy of Sciences, Beijing 100101, People's Republic of China;}
\affiliation{School of Astronomy and Space Science, University of Chinese Academy of Sciences, Beijing 100049, People's Republic of China;}

\author[0000-0003-2471-2363]{Haibo Yuan (苑海波)}
\affiliation{Institute for Frontiers in Astronomy and Astrophysics,
Beijing Normal University, Beijing, 102206, People's Republic of China; }
\affiliation{Department of Astronomy, Beijing Normal University, Beijing 100875, People's Republic of China; }

\author[0000-0002-1027-0990]{Subo Dong (东苏勃)}
\affiliation{Department of Astronomy, School of Physics, Peking University, Yiheyuan Rd. 5, Haidian District, Beijing, China, 100871;}
\affiliation{Kavli Institute of Astronomy and Astrophysics, Peking University, Yiheyuan Rd. 5, Haidian District, Beijing, China, 100871;}

\author[0000-0002-7003-7613]{Chenfa Zheng (郑晨发)}
\affiliation{Department of Astronomy, Beijing Normal University, Beijing 100875, People's Republic of China; }

\author[0009-0000-7775-0297]{Shenzhe Cui (崔深哲)}
\affiliation{Department of Astronomy, Beijing Normal University, Beijing 100875, People's Republic of China; }

\author[0000-0003-1218-8699]{Yi Ren (任逸)}
\affiliation{College of Physics and Electronic Engineering, Qilu Normal University, Jinan 250200, People's Republic of China;}

\author[0009-0002-4542-8046]{Haozhu Fu (付皓竹)}
\affiliation{Department of Astronomy, School of Physics, Peking University, Yiheyuan Rd. 5, Haidian District, Beijing, China, 100871;}
\affiliation{Kavli Institute of Astronomy and Astrophysics, Peking University, Yiheyuan Rd. 5, Haidian District, Beijing, China, 100871;}

\author[0000-0003-3250-2876]{Yang Huang(黄样)}
\affiliation{CAS Key Laboratory of Optical Astronomy, National Astronomical Observatories, Chinese Academy of Sciences, Beijing 100101, People's Republic of China;}
\affiliation{School of Astronomy and Space Science, University of Chinese Academy of Sciences, Beijing 100049, People's Republic of China;}

\author[0000-0002-6790-2397]{Zhou Fan(范舟)}
\affiliation{CAS Key Laboratory of Optical Astronomy, National Astronomical Observatories, Chinese Academy of Sciences, Beijing 100101, People's Republic of China;}
\affiliation{School of Astronomy and Space Science, University of Chinese Academy of Sciences, Beijing 100049, People's Republic of China;}


\begin{abstract}
   Utilizing high-cadence and continuous $g$- and $r$-band data over three nights acquired from the 3.6-meter Canada France Hawaii Telescope (CFHT) aimed to find short-duration microlensing events, we conduct a systematic search for variables, transients, and asteroids across a $\sim1^\circ$ field of view of the Andromeda Galaxy (M\,31). We present a catalog of 5859 variable stars, yielding the most extensive compilation of short-period variable sources of M\,31. We also detected 19 flares, predominantly associated with foreground M dwarfs in the Milky Way. In addition, we discovered 17 previously unknown asteroid candidates, and we subsequently reported them to the Minor Planet Center. Lastly, we report a microlensing event candidate C-ML-1 and present a preliminary analysis.
\end{abstract}



\section{Introduction} \label{sec:intro}

Over the past decades, astronomy has undergone a transformative shift toward a time-domain era driven by the unprecedented wealth of observational data. As observations become deeper, increase in cadences, and achieve higher resolutions, a multitude of variables have been unveiled. Notable findings, such as flares, novae, microlensing events, and binary star systems, have made significant contributions to various astronomical fields. Remarkable strides in time-domain astronomy have been achieved by many projects, such as OGLE (\citealt{1992AcA....42..253U}), ASAS-SN (\citealt{2017PASP..129j4502K}), ZTF (\citealt{2019PASP..131a8002B}) and GWAC (\citealt{2009AIPC.1133...25G}). A number of upcoming time-domain projects, e.g., LSST (\citealt{2002SPIE.4836...10T}), WFST (\citealt{2016SPIE10154E..2AL}), Mephisto (\citealt{2019gage.confE..14L}), SiTian (\citealt{2021AnABC..93..628L}), and Roman  (\citealt{2023arXiv230610647G}), are anticipated to bring about breakthroughs.

As the nearest spiral galaxy, M\,31 presents an ideal celestial laboratory for systematic exploration of diverse variable sources situated at nearly identical distances. Utilizing ultra-wide, high-resolution, sky survey telescopes, variable sources were periodically monitored, ranging from minutes to years, thus significantly enriched our understanding of M\,31's variable source population. In 2004 and 2005, MegaCAM mounted on the Canada France Hawaii Telescope (CFHT), with a 1-degree$^2$ FOV, took images over approximately 50 epochs and detected more than 2,500 Cepheids (\citealt{2012Ap&SS.341...57F}). From 2010 to 2012, the Pan-STARRS 1 (PS1, \citealt{2016arXiv161205560C}) survey gathered around 300 epochs of data and discovered nearly 2,000 Cepheids (\citealt{2013AJ....145..106K}) and approximately 300 eclipsing binaries (\citealt{2014ApJ...797...22L}). The discoveries were later updated with a total of 738 epochs, expanding the catalog to 2,686 Cepheids (both Type I and II) (\citealt{2018AJ....156..130K}, hereafter K18). A large number of variable stars, mostly with long periods larger than 100 days, have been found as by-products of M31 microlensing searches. For example, the POINT-AGAPE survey \citep{2004MNRAS.351.1071A} found 35,414 variabless with the Wide Field Camera on the Isaac Newton Telescope, and the WeCAPP project \citep{2006A&A...445..423F} reported a catalog of 23,781 variables. 
Due to the non-uniform and inter-day samplings, a large number of variables with periods around or shorter than one day remain undiscovered. Even for identified short-period variables, the severely low sampling rate below the Nyquist frequency may cause significant frequency aliases in the reported periods. The observational selection effects and errors introduced by aliasing could greatly limit the science values of short-period variables.

We carried out an M\,31 pixel-lensing (see, e.g., \citealt{1992ApJ...399L..43C}, \citealt{1996ApJ...470..201G}) program using the CFHT MegaCAM camera to search for short-duration microlensing events due to free-floating planets (see \S~3.4 of \citealt{2021ARA&A..59..291Z} for a review). The MegaCAM observations consist of continuous exposures of 5 minutes in CFHT $g$ and $r$ bands during three nights, and there were 231 epochs over a temporal baseline during $\sim7$ days. Such uniformly sampled data with a relatively wide span over the high-frequency range can effectively overcome the selection effects in discovering short-period variables. It also significantly increases the frequency coverage to reach 0.0016 Hz until aliasing occurs, providing a unique opportunity to study short-period variables in M\,31 comprehensively.

The paper is structured as follows. \S \ref{sec:dataset} provides an overview of the observations and data reduction procedures. The process of variable identification is discussed in \S \ref{sec:Variable source detection}. \S \ref{sec:Variable sources classification} introduces various catalogs of the detected variable sources. Finally, the paper concludes with a summary and prospect for future research in \S \ref{sec:Conclusions and outlook}.

\section{Data} \label{sec:dataset}

\subsection{Description of the data}

Our data were acquired through MegaCAM installed on CFHT situated at Mauna Kea. MegaCAM consists of 36-piece 2K $\times$ 4K CCD detectors with a pixel scale is 0.187\arcsec, and its field of view is about 1 square degree. The observations were performed using the CFHT $g$-band ($\lambda \simeq 475$\ {\rm nm}$, \Delta \lambda \simeq 154$ \ {\rm nm}) and CFHT $r$-band ($\lambda \simeq 640$\ {\rm nm}$, \Delta \lambda \simeq 148$\ {\rm nm}) filters. To simplify the polynomial background fitting process and enhance the stability of the subtraction algorithm, the original MegaCAM images were subdivided into 360 1K $\times$ 1K pieces, with each CCD divided into 10 overlapping segments.

The images were acquired on October 24, 28, and 30, 2014\footnote{RunID: 14BS04, PI Name: Subo Dong.}, encompassing a total of 111 CFHT $r$-band images and 120 CFHT $g$-band images, each with an exposure time of 300 seconds. The raw images were first processed  using the Elixir System\footnote{\url{https://www.cfht.hawaii.edu/Instruments/Elixir/}}. The typical limiting magnitude is 24.3 in the CFHT $r$-band and 24.8 in the CFHT $g$-band. Through image stacking, the limiting magnitude could be extended to approximately 27. The saturation magnitude is between 17 and 18 in the two bands. The median DIMM seeing was recorded as 0.5\arcsec in $r$ and 0.55\arcsec in $g$.

\subsection{Photometric and astrometric calibration}
To validate the astrometric and photometric calibrations available in the original CFHT/MegaCAM fits headers, we use comparison stars by cross-matching with the PS1 catalog (\citealt{2016arXiv161205560C}). We select PS1 stars with $g$, $r$ and $i$ magnitudes ranging from 15 to 21 and transform them to CFHT $g$ and $r$ magnitudes with the formula given on the CFHT website\footnote{\url{https://www.cadc-ccda.hia-iha.nrc-cnrc.gc.ca/en/megapipe/docs/filt.html}}. We perform aperture photometry on the comparison stars and calibrate them using the ADU-to-magnitude transformation parameters in the fits headers. We find that they are in sufficiently good agreement with the transformed PS1 magnitudes, and no further adjustment is needed. However, we identify obvious discrepancies upon comparing the equatorial coordinates, and we employ a polynomial fitting approach to make the astrometric corrections (see Appendix A for details).


\begin{figure*}
	\centering
	\includegraphics[width=400pt]{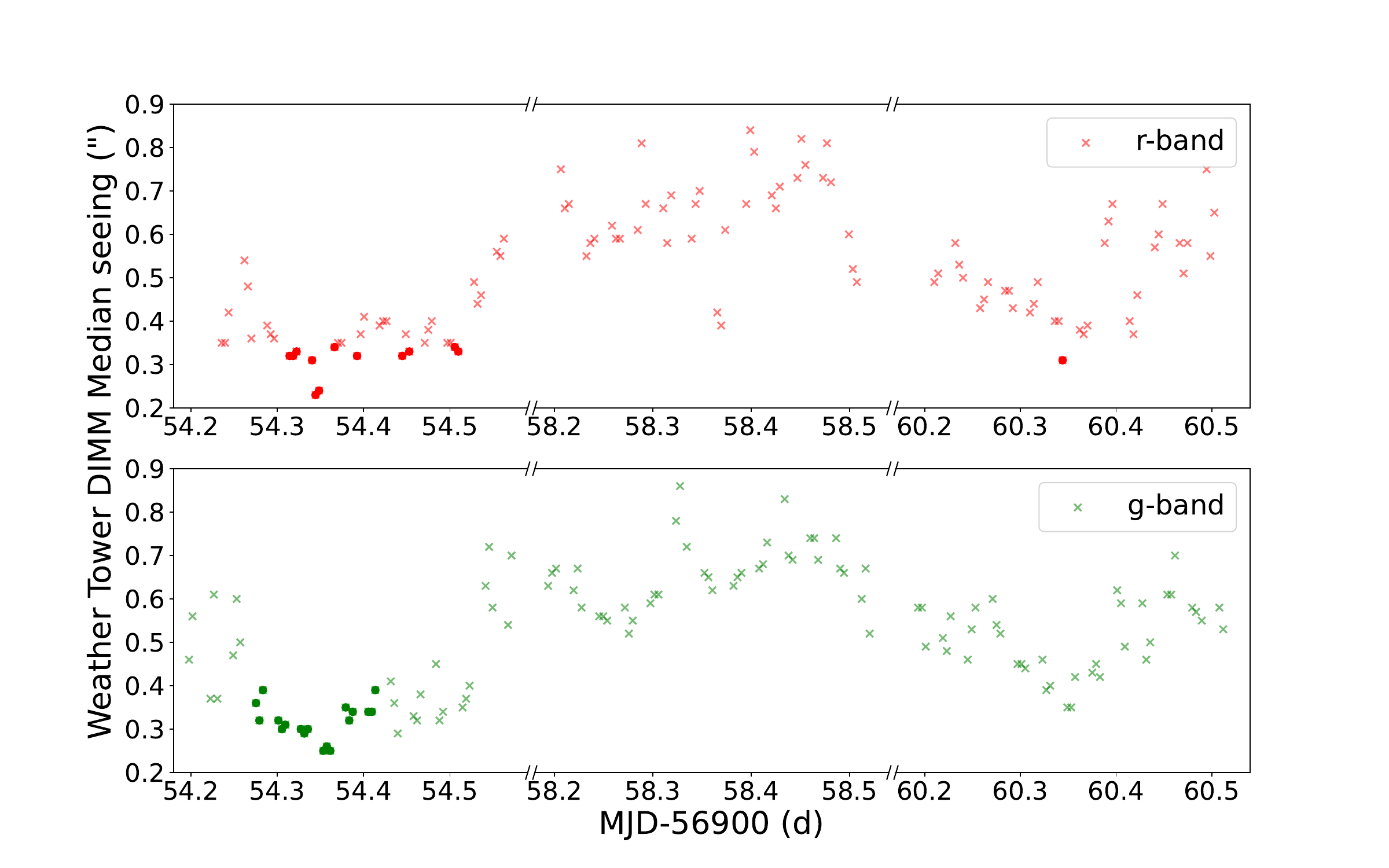}
	\caption{Seeing (red and green crosses for $r$ and $g$ bands, respectively) as a function of time during the three nights of observations. The dots mark those for the reference images.
	}
\end{figure*}

\section{Variable source detection} 
\label{sec:Variable source detection}

In this section, we employ an image subtraction technique to facilitate the identification of variable sources. Following this, catalogs of variable sources are generated for each subtracted image, and subsequent cross-matching of these catalogs is conducted to create an overall catalog of variable sources. By employing statistical analyses and engaging in human visual inspection, we produce the final catalog of variable sources. Subsequently, we proceed to categorize the identified variable sources. The flowchart illustrating this process is presented in Figure 3.

\subsection{Image subtraction} 

 In the crowded regions of M\,31, the use of aperture photometry falls short in ensuring the isolation of photometric targets from the influence of neighboring stars. In pursuit of a precise identification and measurement of variable sources, the employment of the image subtraction algorithm becomes imperative. By subtracting an observed image from the reference image, it effectively eliminates the non-varying sources from the image, thereby facilitating a more robust detection and photometric analysis of variable sources. 
 
 Due to the diverse seeing conditions, even if the luminosity of a star remains constant, the starlight manifests as varied spot sizes on the CCD. This dynamic characteristic renders a direct subtraction approach ineffective, resulting not in a residual background near zero but rather in the presence of a halo. Additionally, the skylight background also varies, which means the subtraction image has a background around a specific value rather than 0. In order to address this intricate challenge, we employ the ISIS subtraction package, an image subtraction algorithm introduced by \cite{1998ApJ...503..325A}. 
 The default configuration file is employed after a number of tests, except that the degree of the polynomial astrometric transform between frames is modified to 2. 

To ensure the effectiveness of the image subtraction process, the reference image should exhibit a higher signal-to-noise ratio (SNR) and a lower full width at half maximum (FWHM) compared to the target images. Additionally, it should be free from cosmic rays and bad pixels. To meet these criteria, we select 18 $g$-band images with 
DIMM seeing less than 0.4\arcsec and 13 $r$-band images with 
DIMM seeing less than 0.35\arcsec (Figure\,1).
These selected images are then stacked as references using a tool within the ISIS subtraction package. The tool transforms all chosen images to one selected image with the same seeing and background level, employing a 3-sigma rejection method. 
Although the FWHM experiences negligible alteration before and after image stacking, there is a marked enhancement in the SNR, accompanied by the elimination of bright pixels from cosmic rays (see Figure\,2).

After the preparation of image subtraction, we execute the ISIS image subtraction procedure on all 1K $\times$ 1K small images for each epoch. Upon inspecting the resulting subtracted images, we find that the subtraction fails for 14\% of the fields (51 failures out of 360 fields) due to various issues and thus exclude them in the following analysis.

\subsection{Sources extraction and photometry}
We first extract the variable source catalog from subtracting each 1K $\times$ 1K small image, followed by performing cross-matching across all catalogs spanning various epochs. Subsequently, targets are selected in the cross-matched catalog, and their corresponding positions are utilized for aperture photometry on the subtraction images, thereby obtaining light curves for each target.

The source extraction process is performed using the DAOStarFinder function from the Python package Photutils. DAOStarFinder employs the same algorithm as the DAOFIND program within the IRAF package.

On the subtracted images, we search for objects showing flux changes with respect to the reference images. As DAOStarFinder expects stars to have positive pixel values on an image, we took the absolute pixel values of each subtracted image before running this routine on them. Besides real sources, these catalog can include bogus sources due to cosmic rays or bad pixels on the CCD. To mitigate these effects, a comprehensive cross-matching procedure across all epochs is requisite. The cross-matching methodology adopted is as follows:

1) The variable source catalog from the initial epoch serves as the benchmark catalog.

2) The variable source catalog from the second epoch is cross-matched with the benchmark catalog from the first epoch using a radius of 3 arcseconds. Sources without any match are incorporated as novel variable sources within the benchmark catalog.

3) The catalog from the third epoch is cross-matched, and the process described in step 2 is iteratively applied.

4) Following the sequential cross-match of variable source candidates from all epochs, we assemble a preliminary variable source catalog encompassing 2,303,513 candidates.

In this candidate catalog, the majority of targets possess a minimum of two positions. Their average positions are employed for aperture photometry in subtracted images, with an aperture radius of 1\arcsec and a sky annulus with an inner radius of 2\arcsec and an outer radius of 3\arcsec. This process enables the generation of light curves with errors for both the CFHT $g$ and $r$ bands measured in ADU. After obtaining the light curve and variability errors in ADU, aperture photometry is performed on the corresponding positions of the reference image with aperture and sky annulus parameters identical to those used previously. To achieve accurate photometry in the dense star fields of M\,31 through aperture photometry, iterative 3-sigma clipping is applied to eliminate stellar pixels in the sky annulus. This process result in a more precise background estimation, enabling a more accurate determination of the brightness of the variable candidates. By adding the previously obtained light curve and its errors in ADU, measured on the subtracted images, to the ADU baseline obtained through aperture photometry on the reference image, we obtain the brightness of each variable  candidate in ADU units for each observation. Utilizing the conversion formula from ADU to magnitudes as specified in the fits header, the light curve aliases is derived.

\subsection{Selection of reliable sources via statistical methods}
We conduct an analysis of the light curves and images  for the selected targets, using several statistical testing methods. It allows us to identify and eliminate targets that are unlikely to be real variable sources.

Within the aperture photometry sky annulus, potential false signals can arise due to factors such as insufficient subtraction of bright stars, misalignment of stars during the subtraction process, and presence of bad pixels on the CCD. While these signals might not be apparent in the light curves, they can be readily identified in the aperture photometry images. To mitigate their impact, we employ the following three criteria for their elimination:

1. The statistical analysis of pixel values within the sky annulus involves the removal of pixels with values exceeding 3 sigma. If the removal ratio for any epoch surpasses an empirical threshold of 16\%, this indicates the presence of bright stars in proximity to the variable source candidate. Consequently, the candidate may not be a genuine variable source and is excluded from the candidate catalog.

2. Measure whether there is a sWtar in the photometry aperture. If a star appears fewer than 6 times in the aperture across all epochs, and these appearances are not in immediate succession (to prevent the inadvertent exclusion of flares), it raises the possibility that the variable source candidate is primarily the result of individual cosmic ray impacts. In such cases, the candidate is eliminated.

3. The presence of a value much lower than the background (100 in our cases) in the photometry aperture suggests the presence of bad pixels. If one or more bad pixels are detected in all epochs, the associated variable source candidate is removed from the catalog.

\begin{figure*}
	\centering
	\includegraphics[width=500pt]{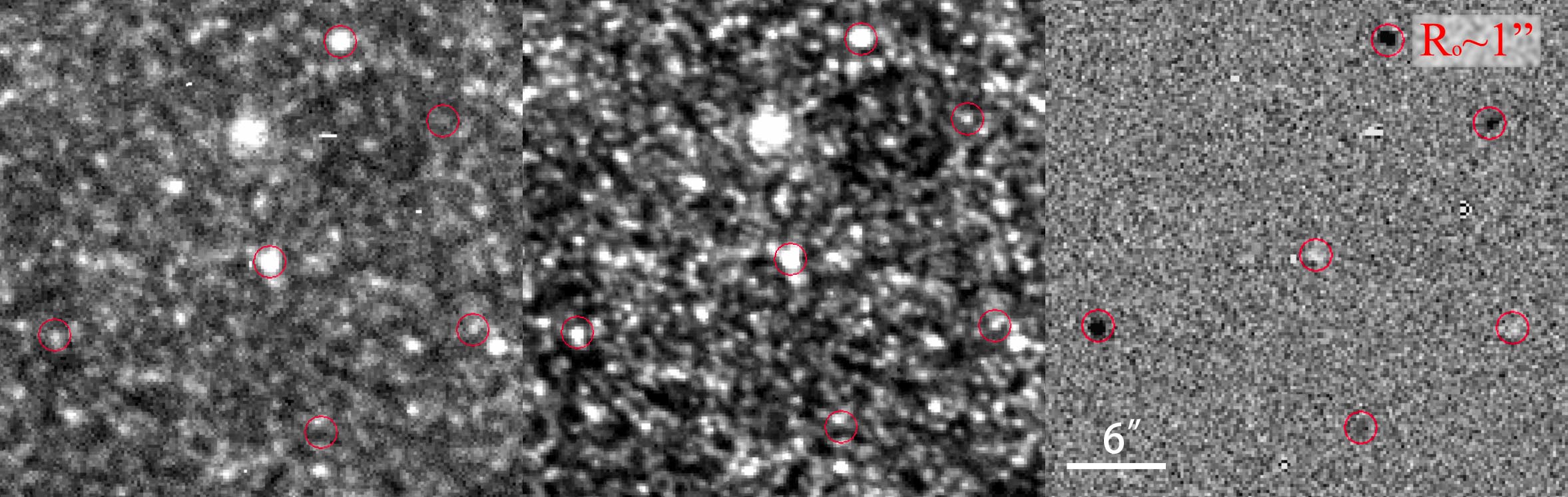}
	\caption{Comparison between a single image (left panel) and its reference image (middle panel) in CFHT $g$-band. Their subtraction image is shown in the right panel. The variable sources in this region are high lighted in the red circles.
	}
\end{figure*}
\begin{figure*}
	\centering
	\includegraphics[width=400pt]{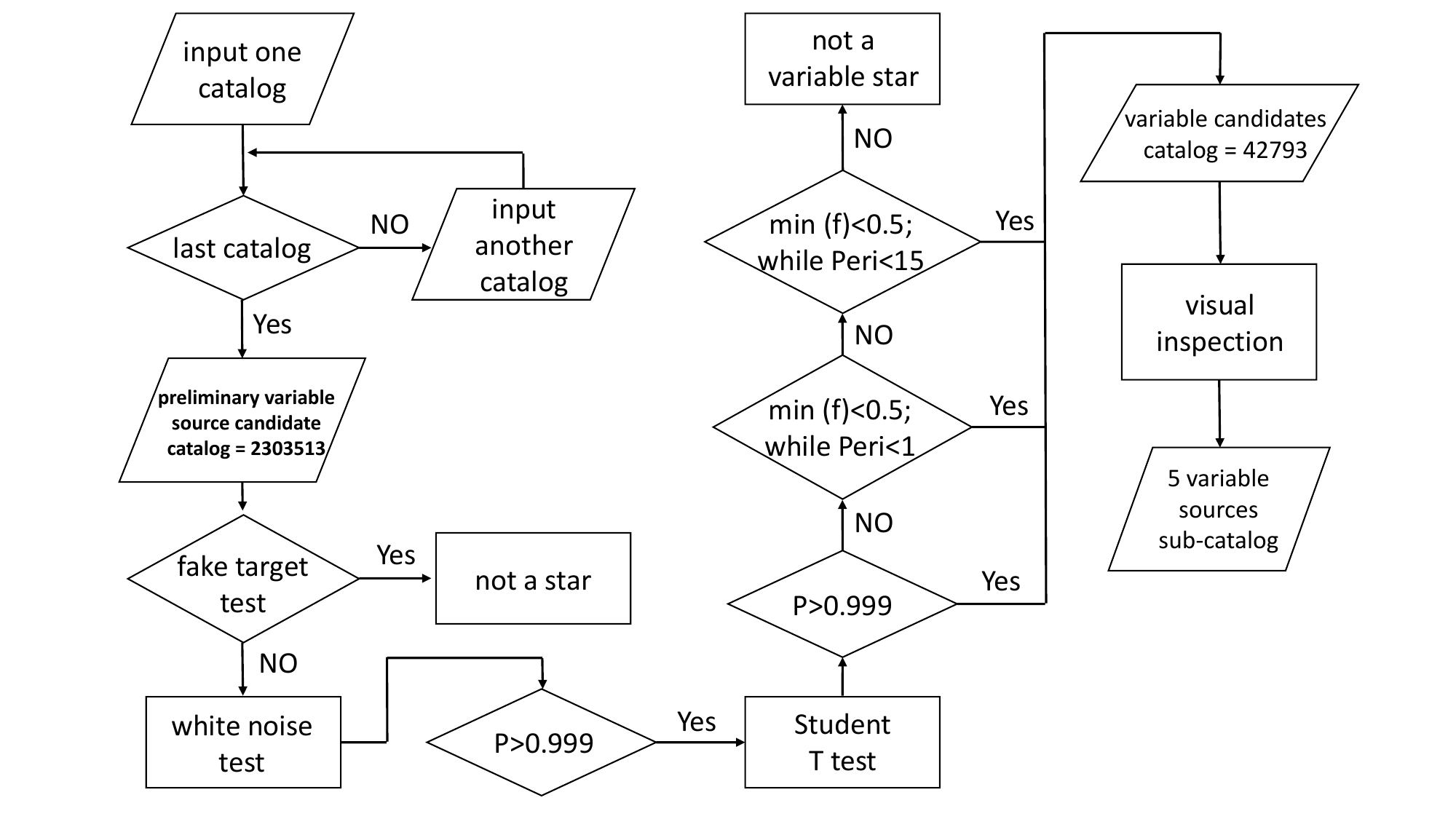}
	\caption{Flowchart of variable source selection. Min (f) here means the minimum period value provided by the PDM algorithm.
	}
\end{figure*}
Following the preceding steps, some spurious sources are effectively filtered out based on the analysis of the images. Subsequently, a statistical testing approach is employed in the following stage to systematically eliminate variable source candidates through the evaluation of their light curves.

Firstly, we perform a white noise test on the light curves. Candidates exhibiting a {\it p}-value of less than 0.999 indicate a distribution consistent with pure noise, devoid of meaningful information. Among all candidates with {\it p}-value exceeding 0.999, we identify two distinct categories: those exhibiting well-defined periodicity and those displaying aperiodic light variations, encompassing phenomena such as microlensing events and flares. To address these different characteristics, we implement two distinct sets of algorithms to identify the two different features of variable sources respectively:

1) The Phase Dispersion Minimization (PDM, \citealt{1978ApJ...224..953S}) algorithm is deployed to identify sources exhibiting periodic variability. We take 0.01 day as the interval to search for the intra-day variable stars with periods ranging from 0.01 to 1 day. For inter-day variability with longer periods, a sampling interval of 0.2 day is adopted. When the period value provided by the PDM algorithm is less than 0.5, it is considered that these variable stars may exhibit the identified period, which will be subject to visual selection to determine whether they are genuine variable stars.

2) To discern potential disparities in brightness across the three distinct nights, we also utilize a Student's t-test. A {\it p}-value exceeding 0.999 indicates a scenario wherein the brightness of the variable source diverges notably between at least one night and the other two nights. Such variations may arise from inter-day variable stars, flares, or microlensing events. These instances are also marked for further scrutiny during the visual selection process.

\subsection{Creating the final variable source catalog via visual inspection}

The above-mentioned automatic procedures result in 42,793 sources. We perform visual inspection to further eliminate spurious sources. For each candidate, we have make a plot to present all relevant information for visual inspection.  An illustrative example of such a visual inspection plot is depicted in Figure 4.

\begin{figure*}
    \centering
    \begin{turn}{-90}
    \begin{minipage}{9in}
    \centering
   \includegraphics[width=650pt]{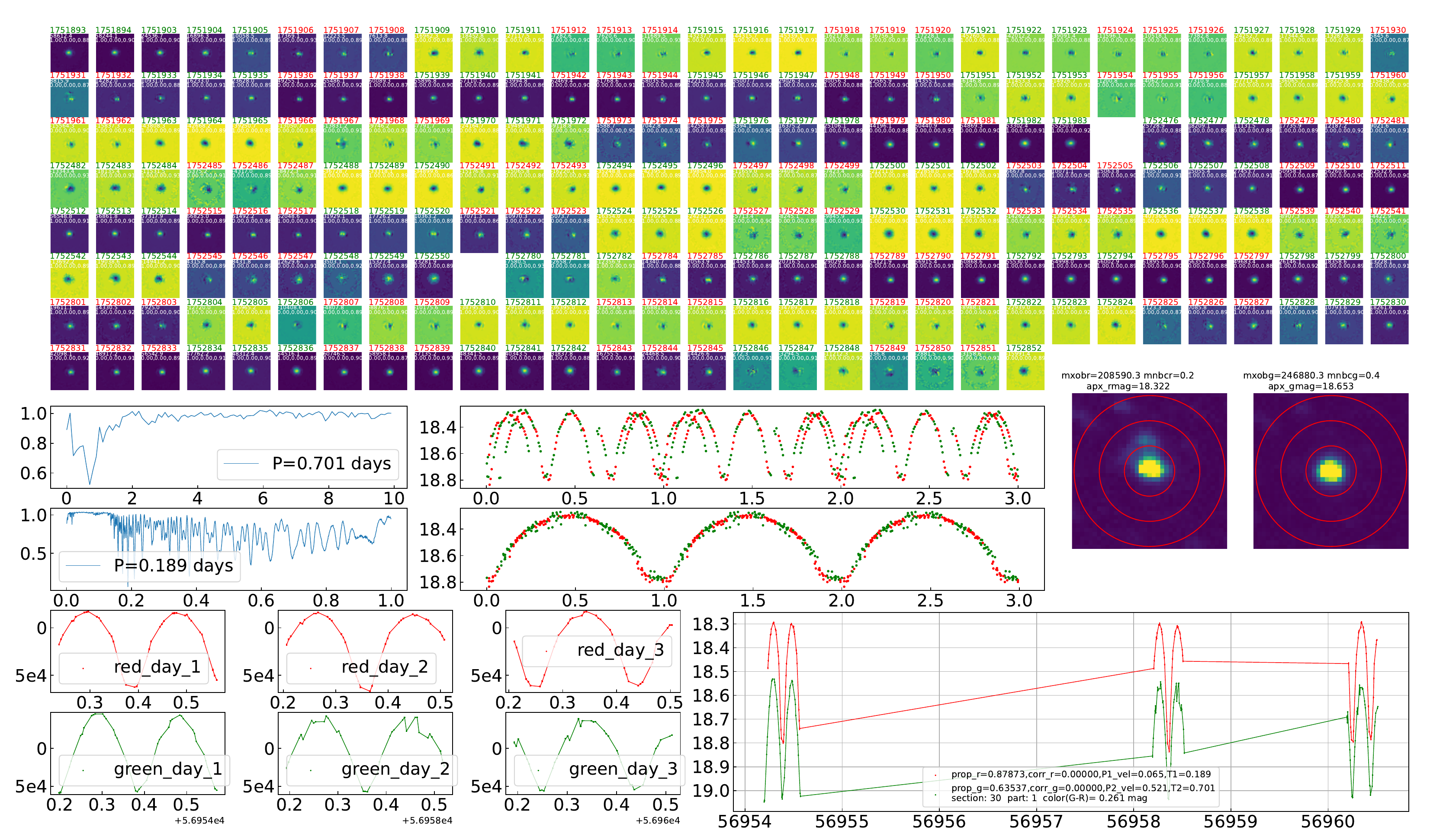}
    \caption{An example plot prepared for each variable candidate for visual inspection. The 231 sub-images on the top represent the portions of subtracted images obtained from a total of 231 observations in the CFHT $g$ and $r$ bands. The numerical colors on the sub-images indicate the observed bands, with green representing the CFHT $g$-band and red representing the CFHT $r$-band. The numbers are sequentially assigned, with smaller numbers indicating earlier observations. The 231 sub-images are divided into three segments, separated by two blank images, representing the three nights of observations. The two large images in the middle-right depict the reference image portions. The four panels in the middle-left represent the period analysis. The left two panels show the relationship between period and intensity obtained using the PDM algorithm, with smaller values indicating more significant periods. The right two panels display folded light curves. The top (bottom) two panels correspond to a search and best-matched folding result for a period range of 0--15 (0--1) days at a resolution of 0.1 (0.001) day. The bottom seven panels depict the time series, with the horizontal axis representing the observation time and the vertical axis representing the intensity, with colors indicating the observed bands. The six individual daily images on the left have the vertical axis represented in ADU values obtained from aperture photometry on the subtracted images, while the large panel on the right has the vertical axis represented in magnitudes. 
	}
    \end{minipage}
    \end{turn}
\end{figure*}

\begin{figure*}
	\centering
	\includegraphics[width=500pt]{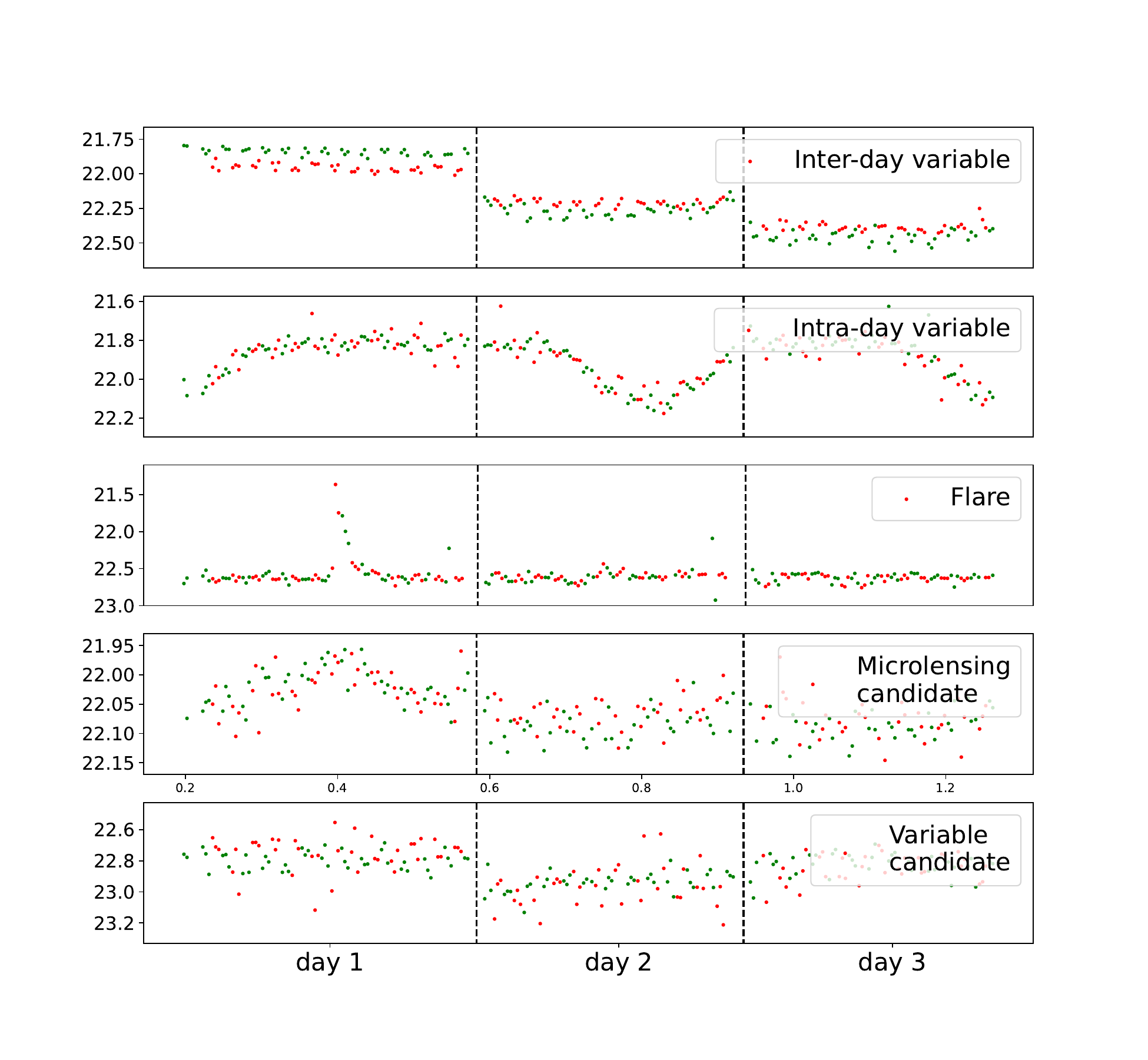}
	\caption{Typical light curves of five different types of variable sources. The upper right legend shows the respective type. The green and red dots represent data in the CFHT $g$ and $r$ bands, respectively. Note that the CFHT $r$ magnitudes have been shifted to a position close to the CFHT $g$ magnitudes by adding the average $g-r$ color. The two vertical dashed lines are used to divide the data into three days.
	}
\end{figure*}

Utilizing visual inspection, we have identified 19 flares, 1200 intra-day variable stars, 3764 inter-day variable stars, 1 microlensing event candidate, and a total of 875 variable source candidates. Examples representing the five distinct variable types are illustrated in Figure 5. 
The criteria employed for visual inspection are outlined as follows:

1) A variable source is categorized as inter-day if a noticeable magnitude disparity exists between observations on different nights, but no significant brightness change can be seen within each individual night. 

2) An intra-day variable classification is assigned to sources displaying significant brightness changes within a single day, such as those exhibiting eclipsing or pulsating behavior. 

3) Rapid increases in brightness observed over a span of several minutes indicate a flare event. 

4) Suspected aperiodic variations, with symmetrical brightening and dimming light curves, are identified as microlensing candidates. 

5) If the computer algorithm described in \S~3.3 classifies a source as a variable star, and there is no error signal caused by observational effects such as cosmic rays and bad pixels, but the amplitude is too small for the human eye to determine whether it is definitely a variable star, it will be classified as a variable source candidate. 

Furthermore, within the candidate variable sources, we also discern variations in brightness induced by asteroids traversing the observed stellar field. Through careful visual inspections and analysis, a total of 24 asteroids have identified across all observation data. For more details on the asteroid detection process, we refer to Appendix B1.

\section{Properties of variable sources} \label{sec:Variable sources classification}

\begin{figure*}[hbtp]
    \centering  
    \includegraphics[width=530pt]{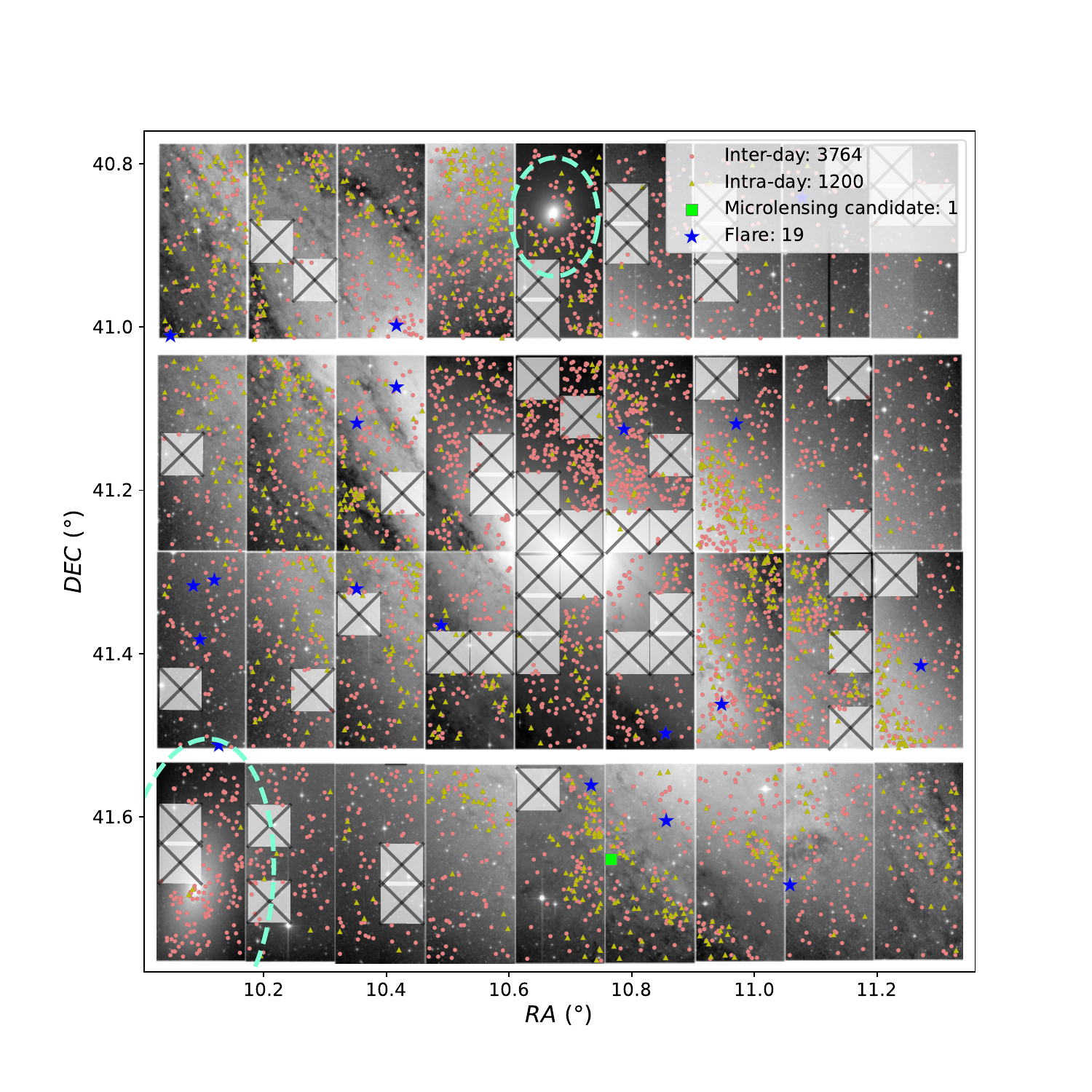}
    \caption {Spatial distribution of different types of variable sources. The black crosses represent the sky area without targets due to  problems in the image subtraction process. The red dots represent the inter-day variable stars, which are evenly distributed in the field of view of M\,31. The yellow triangles represent the intra-day variable stars, which are concentrated in the M\,31 spiral arms. The blue stars indicate the flares. The green square shows the candidate microlensing event. The two cyan ellipses indicate M\,32 and M\,110, whose positions and shape parameters are from \cite{1991rc3..book.....D}. A flag has been added to the variable sources within the two ellipses.
	}
\end{figure*}

Based on the previous section, we have categorized the variable sources into five sub-catalogs: intra-day variables, inter-day variables, flares, variable candidates and microlensing event candidates. The spatial distribution of the first four sub-catalogs is illustrated in Figure 6, while their distribution on the color-magnitude diagram (CMD) is shown in Figure\,7. M\,32 and M\,110, two of M\,31's satellites, are clearly visible in Figure\,6. A flag is assigned to those variables that could be associated with the two satellites. 
The properties of the intra-day and inter-day variable stars seem to be different. Compared to intra-day variables, inter-day variables are redder and fainter, and their spatial distribution is more diffused. After cross-matching with the variable star catalog provided by K18, 
Among the common sources, 97\% of the intra-day variables are identified as type-I Cepheids (FM/FO) in K18. In contrast, 90.5\% of type-II Cepheids (T2) classified in K18 are categorized as inter-day variables. Since type-I Cepheids primarily originate in the galactic arms, and type-II Cepheids are more uniformly distributed throughout the galaxy, this results in distinct spatial distributions between the inter-day and intra-day variables. However, considering that only around 10\% of nearly 6,000 variable sources in this study have common sources with K18, and other long-period variables such as Miras, red supergiants, or even novae in a slow luminosity decline (\citealt{2020MNRAS.496.5503B}) can generate slowly varying light curves and be classified as a inter-day variable, the specific classification of the inter-day variables cannot be precisely determined. This work provides a simple analysis based on the performance of their light curves and the results after cross-matching with the K18 catalog. For the intra-day variables, we have conducted a period check. Due to the limited time span, their light curves may not cover the entire phase or capture critical features. Therefore, intra-day variables of certain periods might remain undetected.

\subsection{Intra-day variables}
Among the entire set of 1200 intra-day variables,  483 of them exhibit detectable periods, including 144 pulsating variable stars with median period of 1.60 days and 189 eclipsing binaries with periods spanning from 0.1 to 4.5 days. Note that for ellipsoidal binary stars or EW-type eclipsing variables that exhibit two identical light curves within one orbital period,  their true orbital periods will be twice the periods provided here. In Figure\,8, we show light curves for 6 pulsating variable stars and 6 eclipsing binaries. The pulsating and eclipsing characteristics can be seen clearly. Additionally, there are 42 variables showing sinusoidal-like light curves. Notably, some of them could potentially be candidates for close binary systems, including intriguing prospects like searching for black hole binaries. The light curves of all 42 such objects are displayed in Figure\,9. Detailed analysis will be presented in future. The remaining subset of 108 stars lacks distinguishing features to be definitively classified.

\begin{figure*}
	\centering
	\includegraphics[scale=0.55]{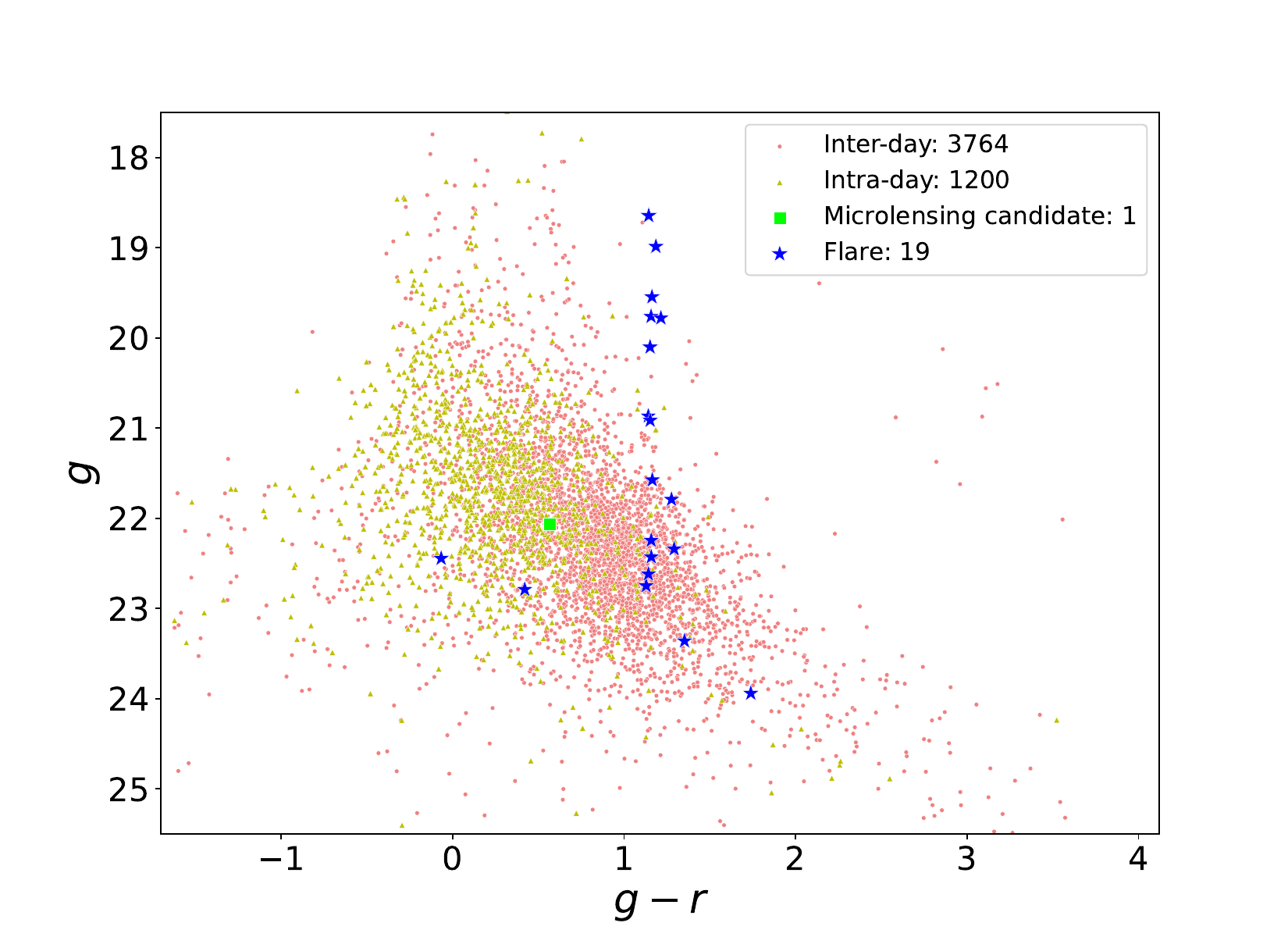}
	\caption{Distribution of different types of variable sources in the color-magnitude diagram.
	}
\end{figure*}

\subsection{Inter-day variable} 

When the duration of variability exhibited by a variable star extends across several tens of days or longer, although significant changes in brightness can be observed over a 7-day span, the variations within a single night may not manifest noticeable. Such sources are classified as inter-day variables, totaling 3764 in our research. The majority of these targets are presumed to be variable stars with periods surpassing 1 day, and such targets have received more comprehensive observations and studies in previous research compared to intra-day variables.

\subsection{Flares} 

Among the 19 flares observed, 16 exhibit $g-r$ colors concentrated around 1, indicating that these are flares on M-type dwarf stars. However, 2 flares were notably peculiar, showing distinct flare shapes but $g-r$ colors are close to 0. Upon examining local images around the flaring stars, we found that these unusual colors were caused by the interference of nearby bright stars and did not represent the true colors of the flare stars. In Figure\,10, we draw the light curves of all 19 flares, aligned to the flare occurrence time.

\subsection{Microlensing event candidate C-ML-1}

We analyze all the suspected microlensing event candidates from visual inspections by fitting their $g-$ and $r-$bands light curves simultaneously using microlensing models. In the standard point-source-point-lens (PSPL) microlensing model \citep{1986ApJ...304....1P}, the magnification function $A(t)$ depends on three parameters, namely, time of the peak $t_0$, impact parameter $u_0$ and the Einstein crossing time $t_{\rm E}$. The flux of a microlensing event scales linearly with magnification as $f(t) = f_s A(t) + f_b$, where $f_s$ and $f_b$ are the fluxes of the microlensed source and the blended light within the PSF, respectively. In the limit of pixel lensing \citep{1996ApJ...470..201G}, where blending is severe ($f_b\gg f_s$) and a detected microlensing event has a high peak magnification $A_{\rm peak}\gg1$ (i.e., $u_0\ll1$), the PSPL model can be well described by two parameters, $t_0$ and the effective timescale $t_{\rm eff} \equiv u_0 t_{\rm E}$. We fit the light curves with both the two-parameter and three-parameter models using Markov chain Monte Carlo (MCMC). We find that the light curves of three suspected candidates are consistent with the PSPL models.

Next, we examine whether a suspected microlensing candidate is physically plausible by placing the source extracted from the microlensing model on the color-magnitude diagram (CMD) of M31. We use the CMD from the deep Hubble Space Telescope (HST) imaging observations of the Panchromatic Hubble Andromeda Treasury (PHAT) survey \citep{2012ApJS..200...18D}. For suspected candidates not falling within PHAT fields, we use a PHAT field at a similar physical distance relative to the M31 center with the target. We convert HST $F475W$ and $F814W$ magnitudes to PS1 $g$ and $r$ magnitudes using common stars in a color-color diagram (Equatios\,1, 2, 3). Subsequently, the PS1 magnitudes are transformed to CFHT $g$ and $r$ magnitudes, utilizing the conversion equations provided on the CFHT web page.\footnote{\url{https://www.cadc-ccda.hia-iha.nrc-cnrc.gc.ca/en/megapipe/docs/filt.html}}.

\begin{equation}
g_{PS1} = F475W-0.067 \times (F475W-F814W)-0.120
\end{equation}
\begin{equation}
r_{PS1} = F475W-0.592 \times (F475W-F814W)-0.097
\end{equation}
\begin{equation}
i_{PS1} = F814W + 0.095 \times (F475W-F814W)+0.327
\end{equation}

For two of the suspected candidates, their source colors derived from microlensing modeling are bluer than all the stars on the respective HST CMDs and thus unphysical. Only one microlensing candidate's source is consistent with the CMD, and we name it C-ML-1 (RA =$00^{\rm h}43^{\rm m}03.86^{\rm s}$, Dec $=41\arcdeg39\arcmin07.3\arcsec$). There are a class of aperiodic variables called ``(blue) bumpers'' \citep{1995ASPC...83..221C, 1997ApJ...486..697A} that can be difficult to distinguish from microlensing events, and they are often blue main-sequence stars (probably Be stars) showing low-amplitude variations. According to \citet{2007A&A...469..387T}, blue bumpers generally exhibit $\sim20\%$ more variations in the red than in the blue. In comparison, the apparent variation amplitude of C-ML-1 in $g$ is $2.3\%\pm1.3\%$ larger than in $r$, which is different from usual blue bumpers.

Figure\,11 shows the light curves and the best-fit PSPL model of C-ML-1. C-ML-1's light curves are equally well fitted by the three-parameter and two-parameter models, with the best-fit $t_0 =  56954.401\pm0.008$ (MJD) and $t_{\rm eff} = 0.11\pm0.03$\,d. Therefore, the source flux is degenerate with the peak magnification in the PSPL model. We set a prior on the impact parameter with a uniform distribution with $0.0001 \le u_0 \le 1$ and run MCMC with the three-parameter PSPL microlensing model. The resulting posterior distributions of C-ML-1 source's color and magnitude are placed on the PHAT CMD using $\sim 800,000$ stars (Figure\,12). The source's color is tightly constrained, while its flux posteriors cover a broad range of stellar types in M31. We plan to carry out further analysis on C-ML-1, including using the theoretical stellar isochrones to constrain the source properties and subsequently microlensing model parameters.


%

\subsection{Variable source candidates} 

Due to the photometric and image subtraction instability in dense regions, variable sources with small amplitudes may only exhibit their variability patterns in certain segments of the light curves or within specific subtracted images in the visual inspection process. To preserve the potential for future investigations of such variable sources, while also preventing confusion with conspicuous variable sources, this study designates these entities as "variable source candidates".

\begin{figure*}
	\centering
	\includegraphics[scale=0.5]{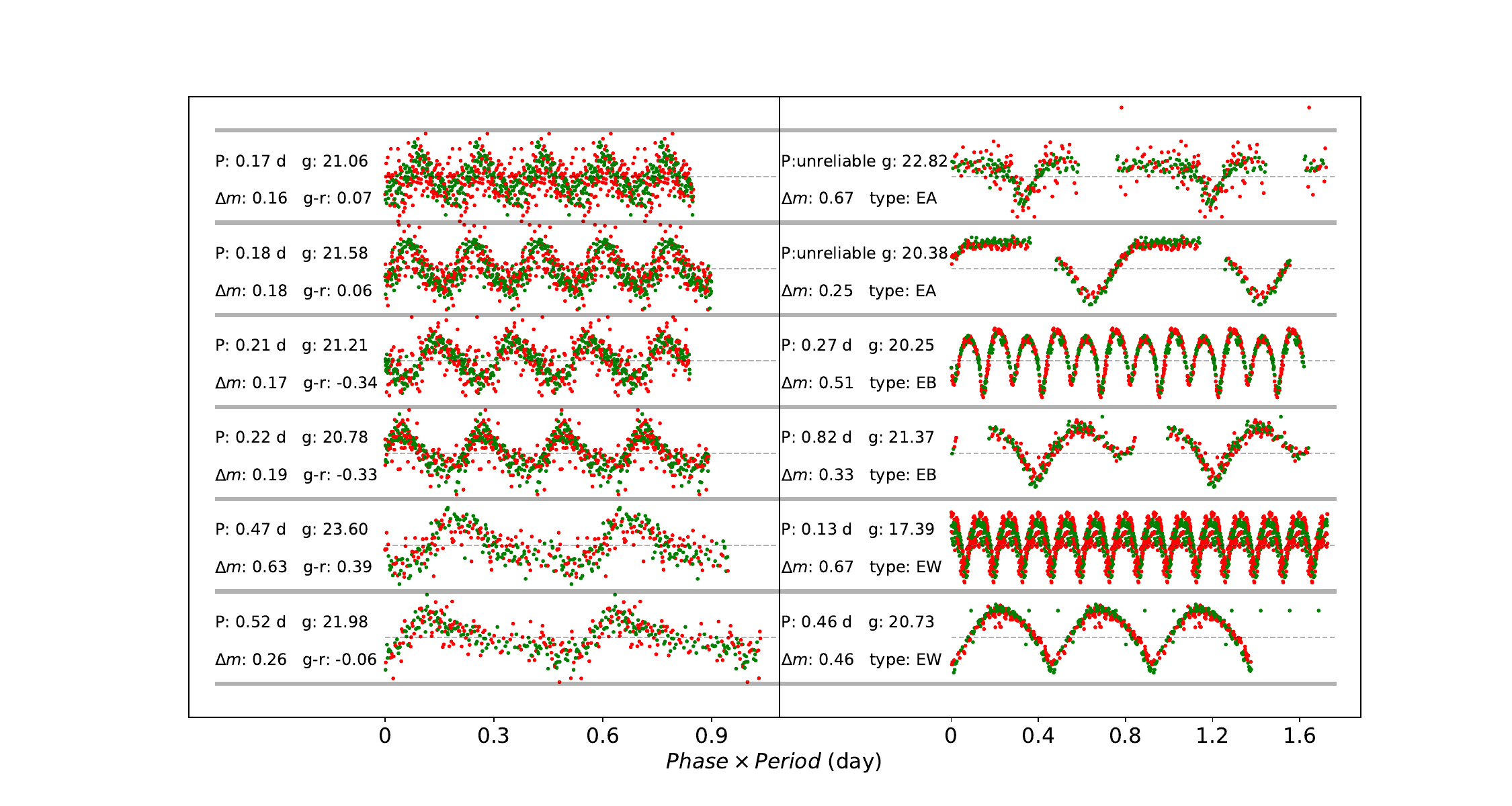}
	\caption{Examples of light curves for pulsating variable stars (left panels) and eclipsing binary stars (right panels).  To the left side of the light curves, some related information is labeled, including  period, CFHT $g$-band magnitude, light variation amplitude,  $g-r$ color or binary type.
	}
\end{figure*}
\begin{figure*}
	\centering
	\includegraphics[scale=0.55]{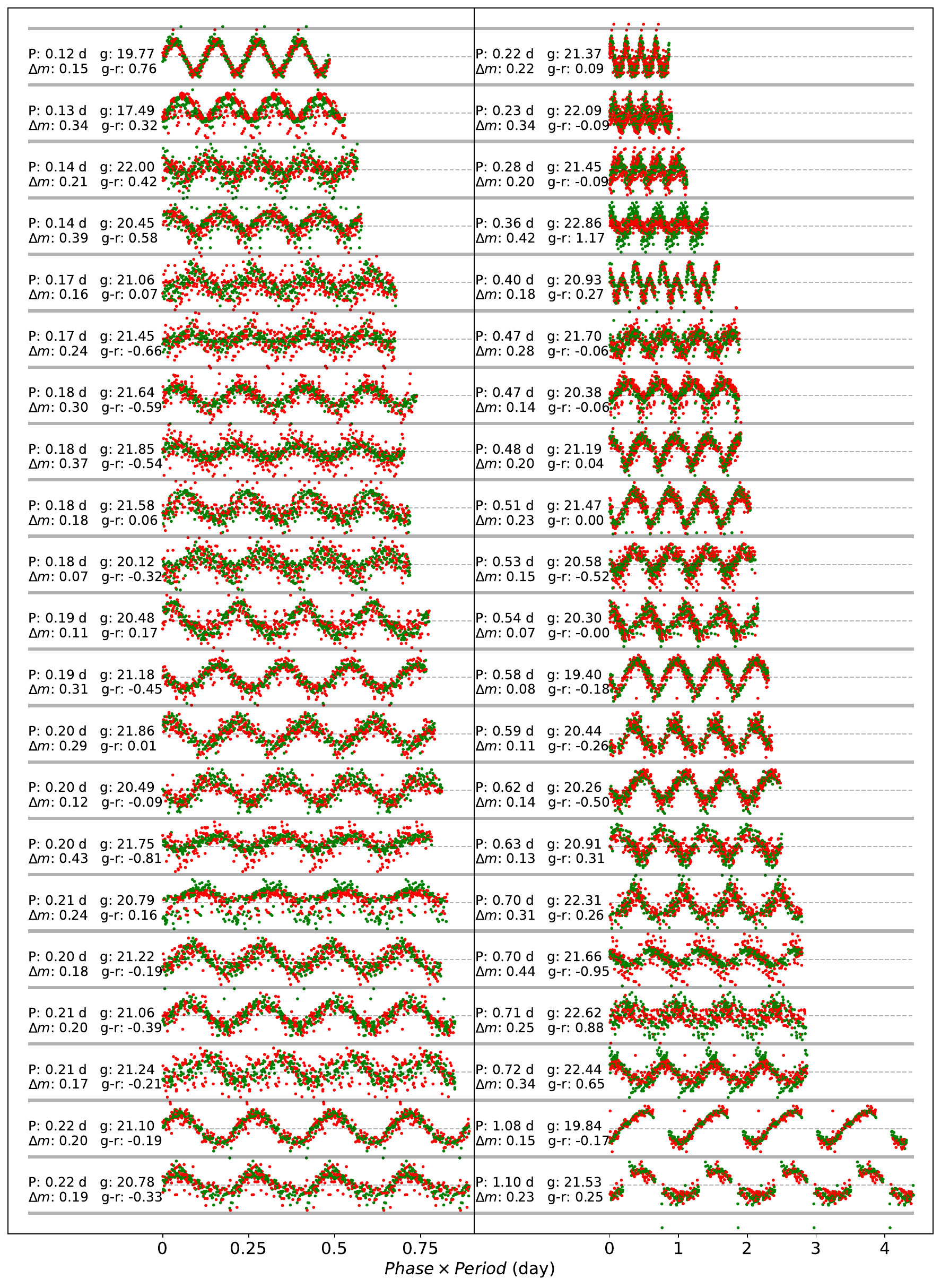}
	\caption{Sinusoidal-like light curves of the 42 variables.  The curved are sorted by their periods.  The period, CFHT $g$-band magnitude, light variation amplitude, and $g-r$ color are labeled.
	}
\end{figure*}
\begin{figure*}
	\centering
	\includegraphics[scale=0.6]{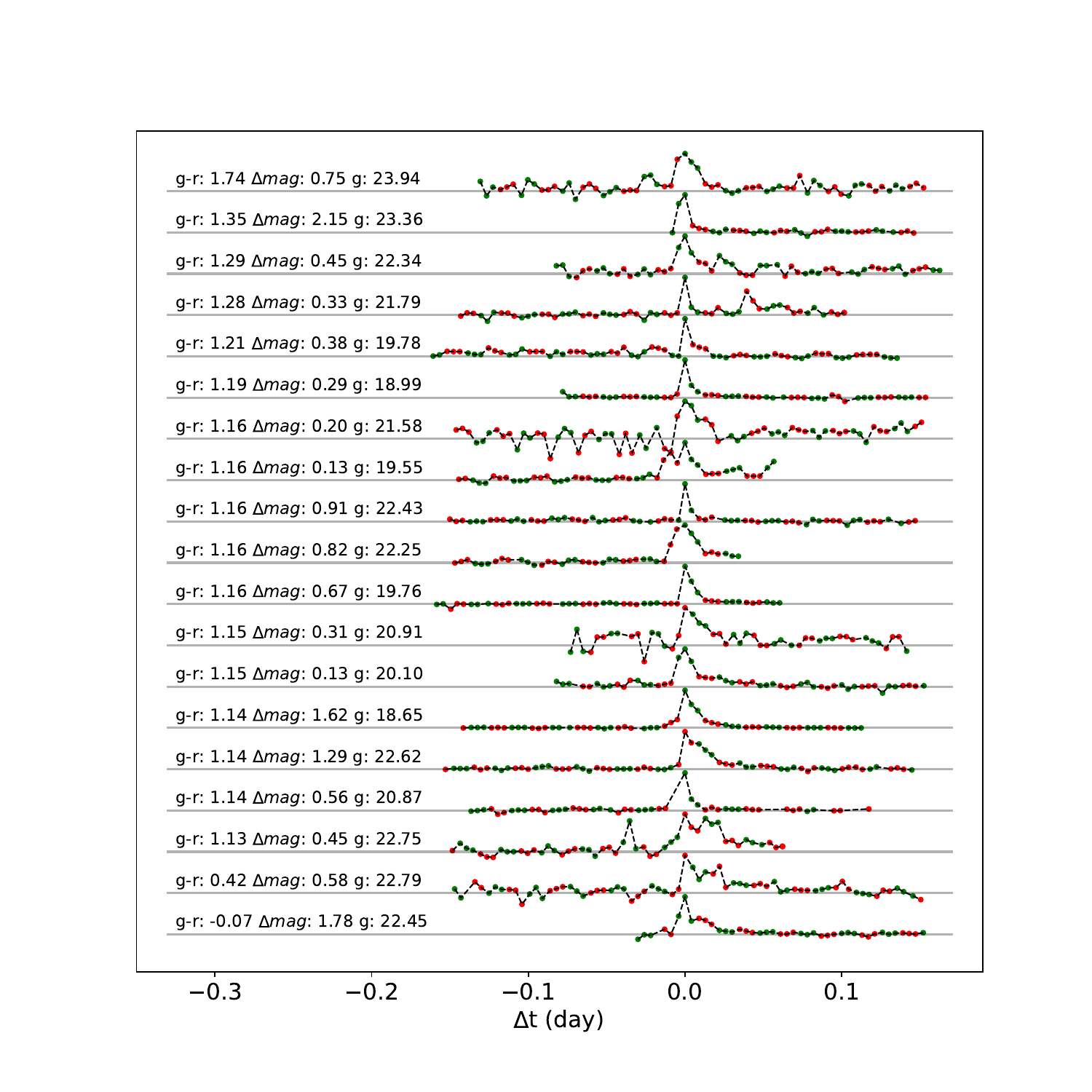}
	\caption{Normalized light curves of the 19 flares.  On the left shows the color, flare amplitude, and magnitude of the stars. The green and red dots represent CFHT $g$ and $r$  magnitudes, respectively. Note the CFHT $r$ magnitudes have been shifted to a position close to the CFHT $g$ magnitudes by adding the average $g-r$ color. 
	}
\end{figure*}

\section{Conclusions} \label{sec:Conclusions and outlook}

\begin{figure*}
	\centering
	\includegraphics[scale=0.3]{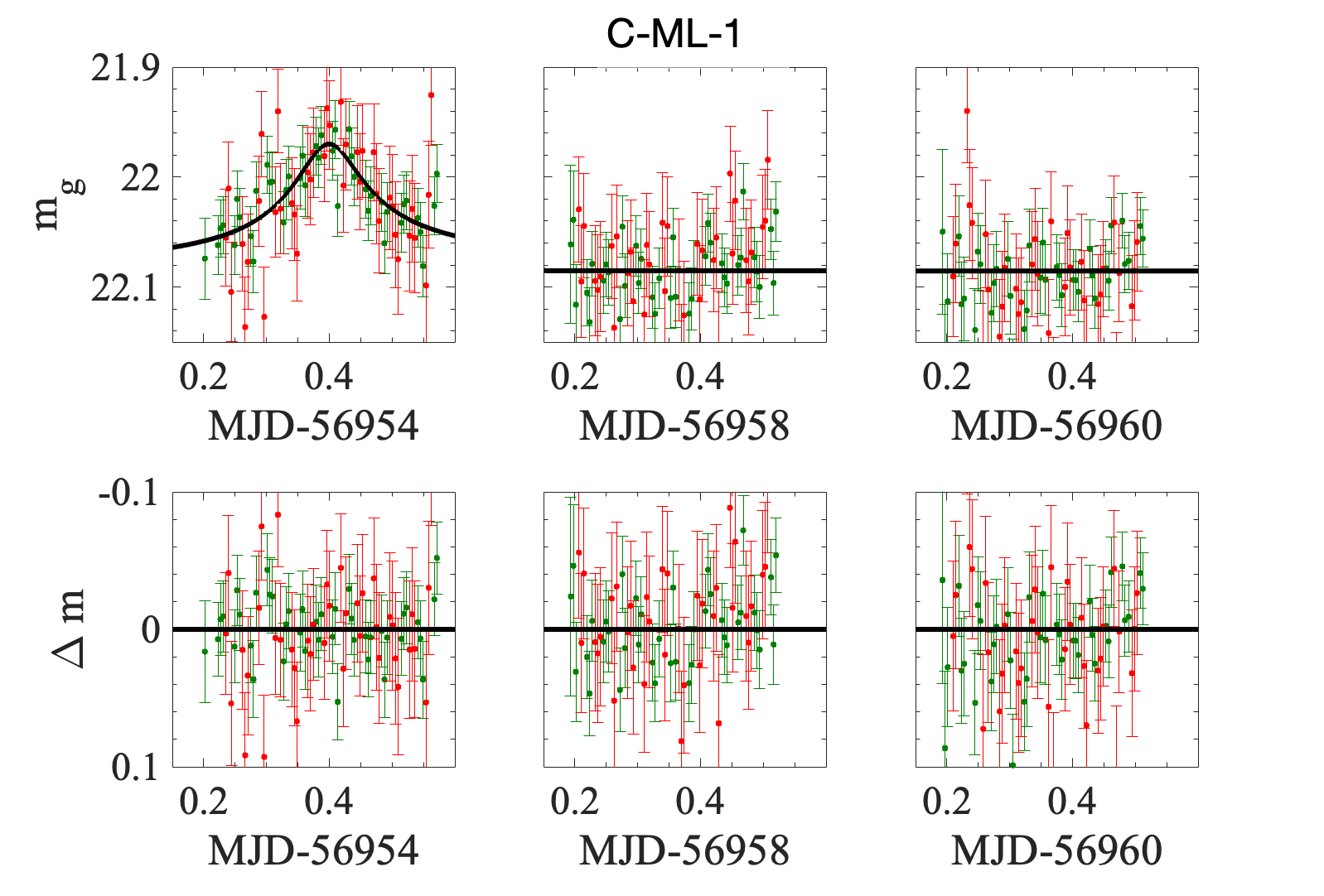}
	\caption{Upper: Light curves of microlensing candidate C-ML-1 observed in $g$ (green) and $r$ (red) shown with the best-fit point-source-point-lens microlensing model (black). The magnitudes are normalized to $g$ band. Lower: The residuals for the microlensing model.}
\end{figure*}
\begin{figure*}
	\centering
	\includegraphics[scale=0.55]{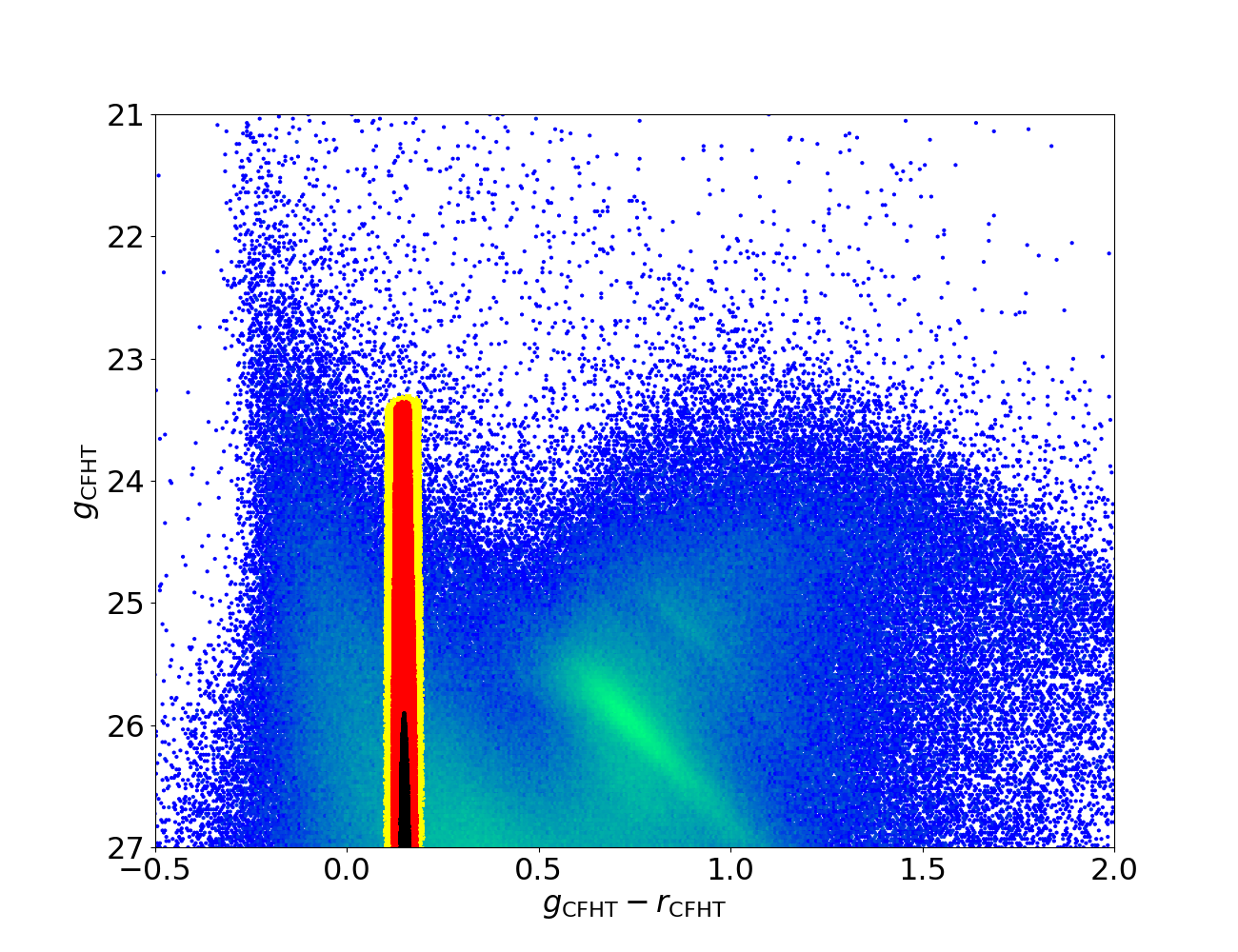}
	\caption{Contours of $\Delta \chi^2=1,4,9$ (black, red, yellow) relative to the minimum of the C-ML-1 source's $g-r$ color and $g$-band brightness from MCMC realizations are shown with the HST color-magnitude diagram (density map) from the PHAT survey. The HST magnitudes are transformed into the CFHT bands. The color and magnitude of C-ML-1's source is consistent with being a star in M31.}
\end{figure*}
\begin{deluxetable*}{ccc}
	\tablecaption{Attributes in the variable catalog. \label{tab:data}}
		\tablehead{	
		\colhead{Attribute} & colhead{Data type}& \colhead{Description}  \label{Tab:data}
		}
	\startdata
		ra\_deg &float& Right ascension (degree) \\
              dec\_deg &float&   Declination (degree)\\
              g\_mean\_mag &float& Mean CFHT $g$ magnitude after 3-sigma clipping \\
              g\_ref\_adu &float&  Flux in the CFHT $g$ band in reference image (ADU)\\
              r\_mean\_mag &float& Mean CFHT $r$ magnitude after 3-sigma clipping \\
              r\_ref\_adu &float&  Flux in the CFHT $r$ band in reference image (ADU)\\
              variable\_type &int&  Five types of variables including 0 for microlensing candidate, 1 for flare, 2 for intra- \\&& day variable, 3 for inter-day variable and 4 for other varible candidates\\
              variable\_subtype &int&  Five types of intra-day variables include 0 for pulsating variable, 1 for pulsating variable candidates, \\&&2 for close binary, 3 for close binary candidate and 4 for sin-like light curve variable \\&&(-9999 means no subtypes was classified)\\
              variable\_subtype\_Period &float& The period of intra-day variable ($-$9999 means undetectable period)\\
              flag & int & 0 for targets in M\,31, 1 for targets might in M\,32, 2 for targets might in M\,110\\
              g\_ref\_img &float 2d array& Star centered sub image in reference image with 37 $\times$ 37 pixels in CFHT $g$ band\\
              g\_obs\_times &float 1d array&  Time of all CFHT $g$ band visits (MJD)\\
              g\_delta\_ADUs &float 1d array&  Flux in the CFHT $g$ band in all subtraction images (ADU)\\
              g\_delta\_ADU\_errs &float 1d array&  Flux error in the CFHT $g$ band in all subtraction images (ADU)\\
              g\_mags &float 1d array&  Magnitude of all CFHT $g$ band visits (mag)\\
              g\_mag\_errs &float 1d array&  Magnitude error of all CFHT $g$ band visits (mag)\\
              r\_ref\_img &float 2d array&  Star centered sub image in reference image with 37 $\times$ 37 pixels in CFHT $r$ band\\
              r\_obs\_times &float 1d array&  Time of all CFHT $r$ band visits (MJD)\\
              r\_delta\_ADUs &float 1d array&  Flux in the CFHT $r$ band in all subtraction images (ADU)\\
              r\_delta\_ADU\_errs &float 1d array& Flux error in the CFHT $r$ band in all subtraction images (ADU) \\
              r\_mags &float 1d array&  Magnitude of all CFHT $r$ band visits (mag)\\
              r\_mag\_errs &float 1d array&  Magnitude error of all CFHT $r$ band visits (mag)\\
	\enddata
\end{deluxetable*}

We have compiled a catalog of distinct variable sources identified within the high cadence CFHT $g$-band and CFHT $r$-band time series data captured by CFHT MegaCAM of the M\,31 galaxy. This catalog encompasses nearly 6000 variable sources, categorized into five classes: inter-day variables, intra-day variables, flares, microlensing event candidates and variable source candidates. With this high cadence observational data, we offer a valuable catalog of variable sources for further explorations and follow up observations, 
including 1) finding potential free-floating planets in M\,31; 2) black hole binaries; 3) intra-day Cepheids; and more. 
The catalog can be found in ChinaVO (\url{https://nadc.china-vo.org/res/r101294/}). 

Additionally, we furnish parameters for 17 newly discovered asteroids, as well as 7 known asteroids. Furthermore, this type of data can also be employed using the same microlensing technique in the search for free-floating planets to constrain the fraction of primordial black holes as dark matter. For instance, \cite{2019NatAs...3..524N} utilized high cadence data of one night taken on November 23, 2014 from Subaru telescope similar to our high cadence data to constrain the fraction of primordial black holes within the mass range of $M_{PBH} \simeq [10^{-11},10^{-6}]M_{\odot}$. By combining our data with theirs, it is possible to delve into the study of variable sources in M\,31 over a longer time baseline. 

\begin{acknowledgments}
{\bf The authors are grateful to the referee for suggestions that improved the quality of the paper significantly.}
This work is supported by the National Natural Science Foundation of China through the projects NSFC 12222301, 12173007, 11603002, 11933004, 12133005, National Key Basic R \& D Program of China via 2019YFA0405500, 2019YFA0405100. We acknowledge the science research grants from the China Manned Space Project with NO. CMS-CSST-2021-A08, CMS-CSST-2021-A09 and CMS-CSST-2021-B12. SD acknowledges the New Cornerstone Science Foundation through the XPLORER PRIZE.
This research uses data obtained through the Telescope Access Program (TAP), which has been funded by the TAP member institutes.

Based on observations obtained with MegaPrime/MegaCam, a joint project of CFHT and CEA/DAPNIA, at the Canada-France-Hawaii Telescope (CFHT) which is operated by the National Research Council (NRC) of Canada, the Institut National des Science de l’Univers of the Centre National de la Recherche Scientifique (CNRS) of France, and the University of Hawaii. 
This work has made use of data from the European Space Agency (ESA) mission {\it Gaia} (\url{https://www.cosmos.esa.int/gaia}), processed by the Gaia Data Processing and Analysis Consortium (DPAC, \url{https:// www.cosmos.esa.int/web/gaia/dpac/ consortium}).
Funding for the DPAC has been provided by national institutions, in particular the institutions participating in the Gaia Multilateral Agreement.

\end{acknowledgments}

\appendix
\section{WCS correction}

The pixel coordinates of stars in the images are converted to celestial coordinates using the WCS information provided in the FITS header. A comparison between the WCS coordinates and the PS1 coordinates reveals a discrepancy, which is subsequently corrected using a third-order polynomial fit of the relationship between the celestial coordinate differences and the pixel coordinates. The upper panels of Figure\,A1 display the coordinate differences for stars utilized in the pixel coordinate correction. The lower panels plot the distribution of error residuals after the correction, demonstrating a notable transition from two irregular clusters to a Gaussian distribution.

\begin{figure*}
	\centering
	\includegraphics[width=480pt]{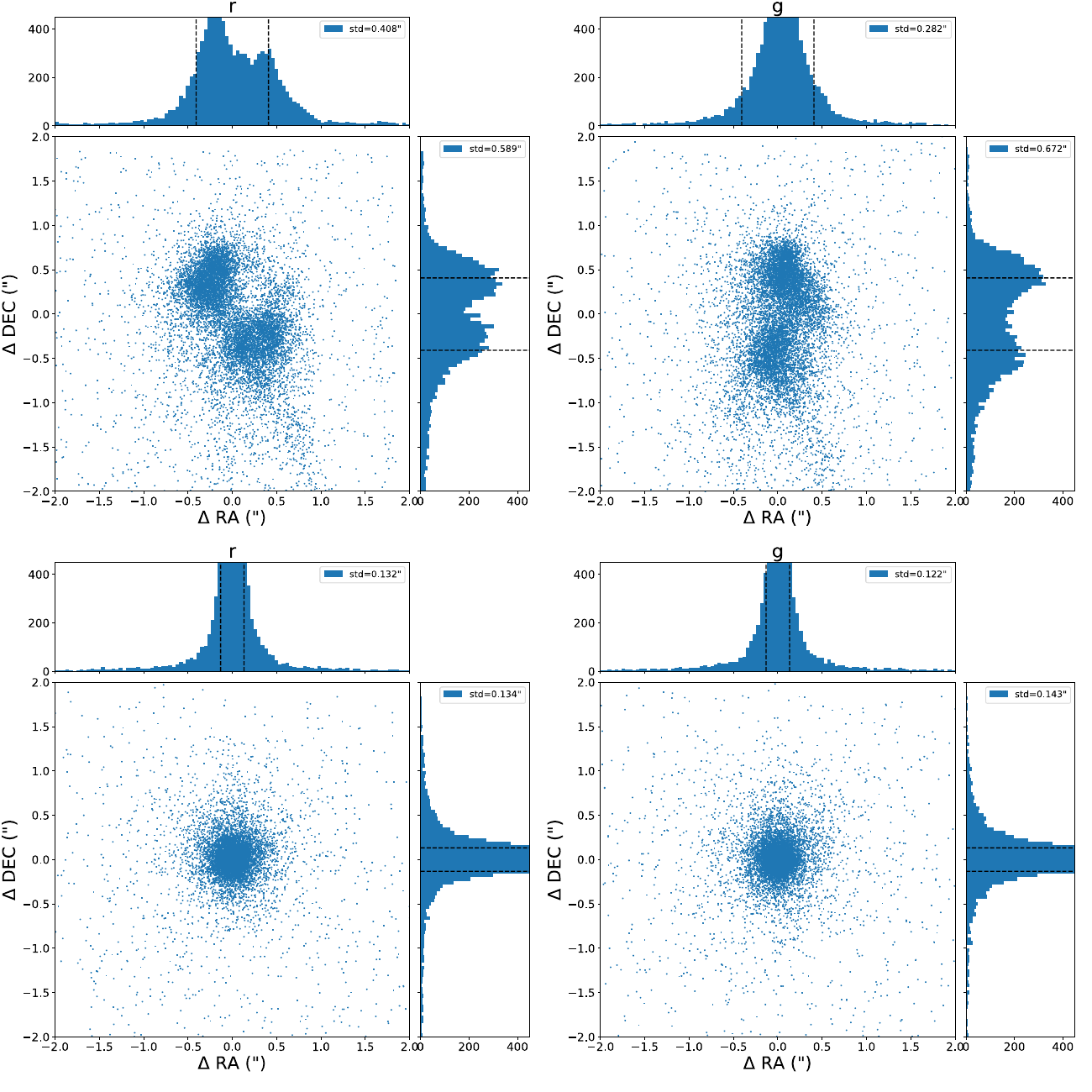}
        \setcounter{figure}{0} 
        \renewcommand\thefigure{\Alph{section}\arabic{figure}}
	\caption{Top panels: error distributions of the original WCS. Bottom panels: error distributions after the correction.}
\end{figure*}

\section{Asteroids}

Through visual inspection, we find that several variable sources  are actually moving objects. After calculate their angular velocities, we classify them as asteroid candidates. However, the efficiency of asteroid detection using visual inspection image is quite low. We then make videos with the time series of M\,31 images to search moving  objects comprehensively.

Among these videos we have found 24 asteroids and the trajectories of these asteroids across the field of view of M\,31 are shown in Figure\,B1. We conduct photometry on the asteroids using aperture photometry and determine their orbital parameters using tools available on the Minor Planet Center (MPC) (\url{https://www.projectpluto.com/fo.htm}). Among them, seven are known asteroids, and the other 17 are newly discovered promising asteroid candidates in this work.

\begin{figure*}
    \centering  
	\includegraphics[width=500pt]{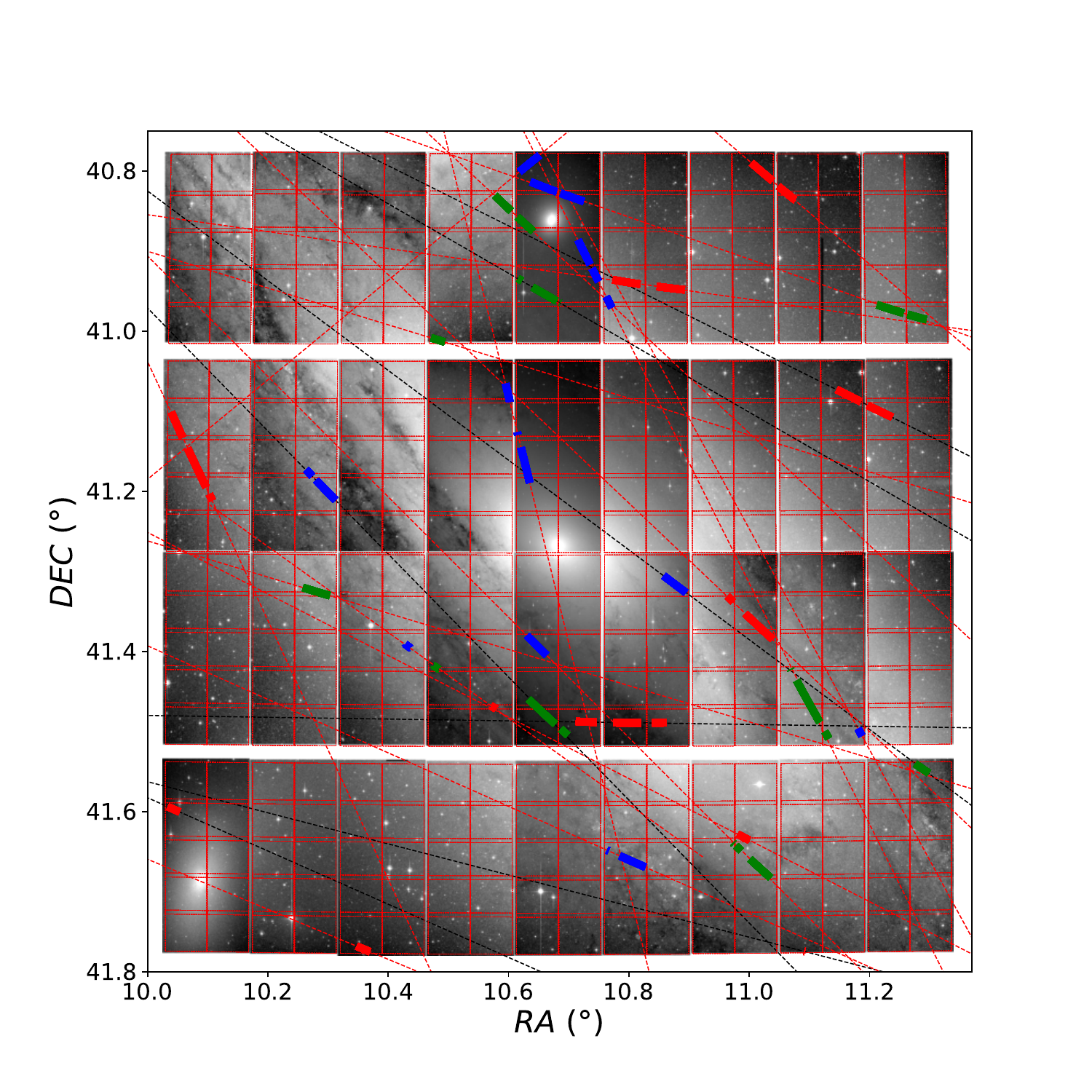}
        \setcounter{figure}{0} 
        \renewcommand\thefigure{\Alph{section}\arabic{figure}}
	\caption {Trajectories of the 24 asteroids. The thick red, green and blue lines represent the paths of the asteroids on three different days. The thin red dotted lines indicate new asteroid candidates we discovered, and the thin black dotted lines indicates known asteroid. 
	}
\end{figure*}
\newpage
\bibliography{paper.bib}{}
\bibliographystyle{aasjournal}



\end{CJK*}
\end{document}